\documentclass[12pt,fleqn]{article}
\usepackage{xiiiemc}
\usepackage{natbib}
\usepackage{fancyhdr}
\usepackage{fancyvrb}
\usepackage{color}
\usepackage{wallpaper} 
\usepackage{titlesec}   %% Define space between paragraph e section
\usepackage{float} 	%% Use to fix Figure or Table: ex: \begin{table}[H]
\usepackage[usenames,dvipsnames]{xcolor}
\usepackage{hyperref} 	%% Use to fix Figure or Table: ex: \begin{table}[H]
\hypersetup{
	pdfcreator={LaTeX with abnTeX2},
	pdfkeywords={abnt}{latex}{abntex}{USPSC}{trabalho acadêmico}, 
	colorlinks=true,       		% false: boxed links; true: colored links
	linkcolor=blue,          	% color of internal links
	citecolor=blue,        		% color of links to bibliography
	filecolor=magenta,      		% color of file links
	urlcolor=blue,
	allbordercolors=black,
	bookmarksdepth=4
}
\usepackage{tocloft}
\titleformat{\section}
  {\normalfont\bfseries}{\thesection.}{0.5em}{}
 
\renewcommand\thesection{\arabic{section}}

\let\OLDthebibliography\thebibliography
\renewcommand\thebibliography[1]{\OLDthebibliography{#1} \setlength{\parskip}{0pt}\setlength{\itemsep}{0pt plus 0.3ex}}

\usepackage{grffile}
\usepackage{afterpage}

	\title{THE ALGORITHMIC-AUTOREGULATION (AA) METHODOLOGY AND SOFTWARE: A COLLECTIVE FOCUS ON SELF-TRANSPARENCY}

\author
    {\rm \begin{tabular}{l} 
    \textbf{Renato Fabbri}$$ - {\textnormal renato.fabbri@gmail.com}\\%
    {\fontsize{11}{0}\selectfont University of São Paulo, Institute of Mathematical and Computer Sciences - São Carlos, SP, Brazil}\vspace*{-0.05cm} \\
  \end{tabular}}
\fancypagestyle{firspagetstyle}
{
	\lhead{}
	\fancyhead[C]{%
		\includegraphics[width=0.9\linewidth]{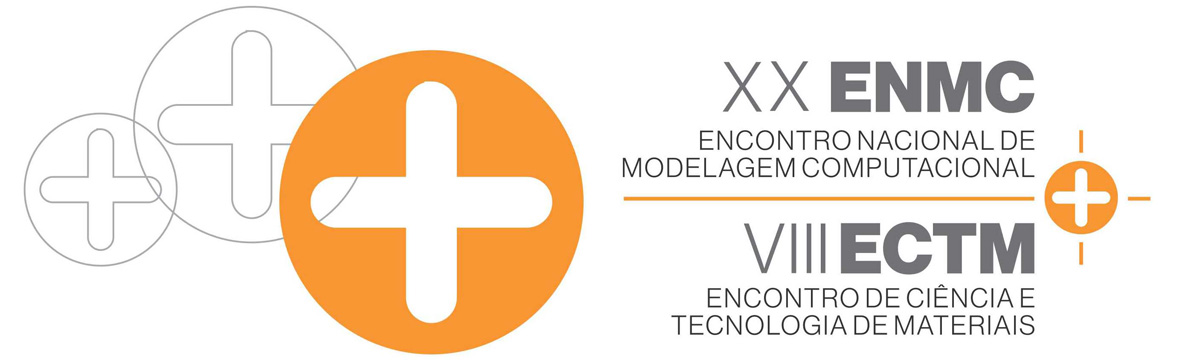}\\%
		{\scriptsize \fontfamily{phv}\fontseries{b}\selectfont \color[rgb]{0.45,0.45,0.45}
		16 a 19 de Outubro de 2017\\
		Instituto Politécnico - Universidade do Estado de Rio de Janeiro\\
		Nova Friburgo - RJ\\
	    }
	}
	\renewcommand{\headrulewidth}{0.0pt}
	\fancyfoot[C]{\footnotesize \parbox{15cm} {\centering  \fontsize{7.5}{0}\selectfont \it Anais do XX ENMC – Encontro Nacional de Modelagem Computacional e VIII ECTM – Encontro de Ciências e Tecnologia de Materiais,  Nova Friburgo, RJ – 16 a 19 Outubro 2017}} % \ttfamil
	\rhead{}
}

\begin{document}
\maketitle

\thispagestyle{firspagetstyle}

\fancyhead[L]{\footnotesize{\fontsize{7.5}{0}\selectfont \it XX ENMC e VIII ECTM\\
	16 a 19 de Outubro de 2017\\
	Instituto Politécnico Universidade do Estado do Rio de Janeiro – Nova Friburgo - RJ\\}}
\renewcommand{\headrulewidth}{0.0pt}
\fancyfoot[C]{\footnotesize \parbox{15cm} {\centering  \fontsize{7.5}{0}\selectfont \it Anais do XX ENMC – Encontro Nacional de Modelagem Computacional e VIII ECTM – Encontro de Ciências e Tecnologia de Materiais,  Nova Friburgo, RJ – 16 a 19 Outubro 2017}} % \ttfamil
\rhead{}

\begin{abstract}
There are numerous efforts to achieve a lightweight and systematic account of what is done by a group and its individuals.
The Algorithmic-Autoregulation (AA) is a special case,
in which a technical community embraced the challenge of registering their own dedication
for sharing processes, self-transparency, and documenting the efforts.
AA is used since June/2011 by dozens of researchers and software developers,
with the support of different software gadgets and for distinct tasks.
This article describes these implementations and statistics of their usage
including expected natural properties and ontological formalisms 
which eases comparative analysis and furthers integration.
\end{abstract}


\pagestyle{fancy}

\section{INTRODUCTION}\label{sec:intro}\label{sec:start}
The Algorithmic Autoregulation (AA) is a self-transparency mechanism for sharing processes,
documenting efforts, and enhancing personal or collective self-transparency.
The purposes for using AA are numerous:
enabling automated and fair compensation for dedications,
facilitating co-working, introducing newcomers,
keeping public historical logs of activities, etc.
Indeed, other systems have been designed for such a task (see Section~\ref{sec:rel}).
A brief characterization of AA is:
\begin{itemize}
    \item The collective origin, purpose and maintenance.
	    This is a free-culture trait, present within many software,
		and leads to open software and data as described in Section~\ref{sec:sofsup}.
    \item Voluntary logging of messages about ongoing work.
    \item Enables coordinating distributed team work through individual merit.
    \item More a practice than a software: AA presents variations on the software support and message composition.
	    Often features found are screencasts, peer validation and periodic messaging.
\end{itemize}

Transparency in this context should be understood as it is
in the State and in enterprises:
a public account of activities~\citep{stso};
and not directly as transparency in self-knowledge,
as is the case in some philosophical and political contexts~\citep{stph}.
One should read~\cite{paaper} for a noteworthy overview of AA as a Global Software Development (GSD) undertake.

\subsection{Related work}\label{sec:rel}
Authors know of no \emph{civil society transparency} platform.
There is a number of transparency initiatives for governments~\citep{govTr},
for religious parties~\citep{espTr} and for private institutions~\citep{priTr}.
The data analysis performed in this document
comprises methods that are derived from Text Mining (TM)
and Complex Networks (CN) constituting a hybrid framework of
approaches that are somewhat established~\citep{cla1} and novel~\citep{nov1,nov2}.

\subsection{Historical note}
In June the 7th, 2011, Cleodon Silva~\citep{cleodon} died by heart failure.
In his memory, the Lab Macambira (LM) group was born (Pedro Macambira was one of this pseudonyms).
The AA was conceived as the ``cardiac pulse'' of the group and is in constant usage since July, 2011.
It gathers thousands of messages, tenths of users and hundreds of processes.
AA messages present contributions, such as commits to the official repositories of Evince,
Firefox, Libreoffice, Puredata, and other software~\citep{paaper}.
A number of other activities were registered:
new software elaboration and coding, writing of articles,
Wikis and Etherpads; articulation of civil society, academic and state instances;
studies and reviews.
Even so, AA is highly biased towards software development,
as can be observed in Section~\ref{sec:stats},~\cite{ensaaio},
and in the GSD article about AA~\citep{paaper}.

\subsection{Essay structure}\label{sec:ess}
Section~\ref{sec:des} describes AA design.
Section~\ref{sec:use} describes AA uses incident and envisioned.
Section~\ref{sec:sofsup} describes different software written or used for AA.
Section~\ref{sec:data} is dedicated to describing the structure of AA data.
Section~\ref{sec:stats} presents statistics about AA in terms of vocabulary and networks.
Section~\ref{sec:conc} concludes with final remarks,
further works, and acknowledgements.

\section{AA DESIGN}\label{sec:des}
To understand use practices and software support (Sections~\ref{sec:use} and~\ref{sec:sofsup}),
one needs to grasp the core design features of AA:
\begin{itemize}
    \item evenly spaced messages should be sent by the AA user.
	    The time lapse is called a ``slot'' and the message a ``shout''.
		A slot might also mean the time lapse and the message,
		this is context dependent and will be disambiguated if needed.
    \item Shouts should report the task being tackled and/or a briefing of what was done in the slot.
    \item Shouts might be grouped into ``sessions''.
	    Each session is ideally linked to a short screencast by the user,
		with a few dozens of seconds for the explanation about the session.
    \item Ideally, each session is sent by email to a random AA user for validation.
    \item Such systematics should be adapted as needed both to generate relevant documentation
	    and to assist the user to keep track of the outcomes of his own dedications.
\end{itemize}

\begin{figure}[!htbp] %h or !htbp
\vspace{-2pt}
\begin{center}
    \includegraphics[width=.7\textwidth]{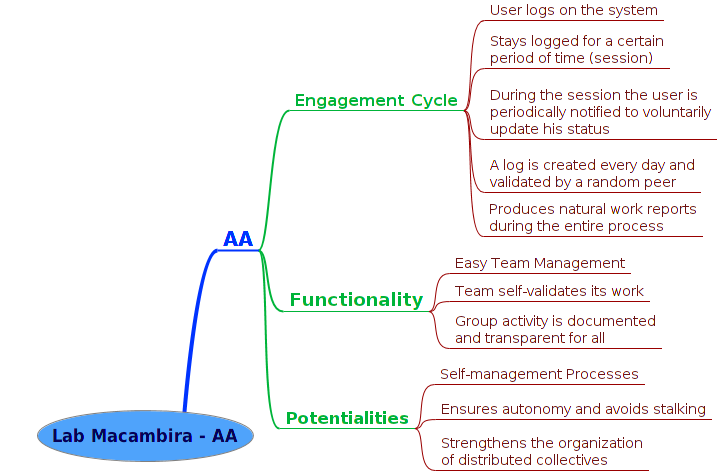}
    \caption{A mind map of the AA methodology shared by users:
	i) Engagement cycle – the usage of AA;
	ii) Functionality – the design goals of the system;
	iii) Potentialities – envisioned benefits of AA by authors of the diagram.
	As seen in Section~\ref{sec:start},
	core benefits emanate from the self-transparency aspect of AA,
	with worthy mentions to testifying dedications and sharing processes.} 
    \label{fig:consult}
\end{center}
\end{figure}

Variants of this features were conceived and used.
Figure~\ref{fig:consult} presents a diagram shared and referenced by AA
users in the first months of AA practice.

\section{USE PRACTICES}\label{sec:use}
Distinct ways to use AA are incident,
mostly regarding the design exposed in Section~\ref{sec:des}.
Even those cases which are far from standard can be understood through the angle of AA paradigm.
Deviations from the ideal case are always present (see Section~\ref{sec:devia}).

\subsection{Words and tags}\label{sec:usewt}
Throughout AA usage, particular words and tags have been used to classify shouts.
Of particular interest are:
\begin{itemize}
    \item Hashtags, such as \#aa, \#coding and \#articulation.
	    These were inherited from Twitter practice. 
    \item Tags starting with ``+'' sign, such as +django, +sna and +reading.
	    These aimed at a particular usage of tagging within AA,
		with independence of other systems and easing concurrent use of AA and other social networks.
    \item Words and abbreviations.
	    Sometimes used in the beginning of shouts,
		others on the end of them,
		these also had the purpose of facilitating the categorization of shouts.
		These cases were sometimes described as tags for entire sessions 
		or for all shouts since tagging, until another tagging was sent by the user (in a shout).
\end{itemize}

These tagging schemes were also used as a way to enable the ``ubiquitous AA'',
i.e. usage of AA in any social network or communication protocol.
The \#aao0 tag was used for shouting within Twitter messages and was the most prominent manifestation of the ubiquitous AA.
Facebook tagging was also used to indicate posts and comments that were shouts.
On some extreme cases, tagging was used in any platform, considering ubiquitous AA implemented,
but not yet mined.

\subsection{Messages (including shouts)}\label{sec:usesh}
Messages for AA usage can be of various types.
Usually, the type was dictated by the first word of the message.
Start messages started a session, while stop messages finished an ongoing session.
Push messages sent local sessions (or independent shouts) to a shared database.
There was only one automatic message, designated to register a ``lost timeslot''
of a session (see Section~\ref{sec:usess}).
Additional messages were used to query for tickets attributed to the user,
milestones and other traditional software development managements facilities.

By far the most important AA related message to date is the ``shout''.
Dedicated to expose ongoing tasks, shouts are recurrently envisioned as a structured message,
in which the user classifies the shout through special words and tags,
and describes ongoing efforts with natural language (usually Portuguese or English).
Examples of structured shout proposals are in~\cite{aaWiki} and~\cite{aaREADME}.
Nevertheless, shouts are used by all AA users, in almost all cases, without such sophisticated structure,
but as a plain short natural language description of current efforts,
i.e. without classification whatsoever of the message. 

\subsection{Sessions}\label{sec:usess}
AA sessions are collections of shouts.
These conventions have had incidence in practice:
\begin{itemize}
    \item Shouts within a session are input by the user each 15 minutes.
    \item Considering the 15 minutes grid, the tolerance for shouts in an ideal session is of $\pm 5$ minutes.
    \item Total duration of 2h per session, in which 8 shouts should outline tasks and technologies.
    \item A very short screencast is recorded at the end of each session,
	    in which the user described the dedication within dozens of seconds.
    \item The session is sent to a random user for peer validation 
	    in which the session receives a score based on the shouts and the screencast.
\end{itemize}

Such a session design was very important in the first 6 months of AA,
where each of the  $\pm 10$ apprentices were performing a session per day.
Other users also delivered sessions, but not as regularly.
Noteworthy: shouts can be separated by durations different from 15 minutes: 
example of incident shout separations include 5 minutes, 2 minutes, 30 minutes, 1h.
Most shouts are not explicitly related to sessions
but still occur in a session-like context.
This is regarded as a consequence of the intuitive usage
sessions were aimed for, and as an inheritance of early AA practice.
Thus, the term ``session'' is used as meaning both a session registered as such,
and as an arbitrary time-contiguous set of shouts from the same user.

\subsection{Accomplishments}\label{sec:usedev}
Processes registered by AA usage often purposes to accomplish something: 
write a software or an article;
make images, music or research scripts; 
articulate groups, read technical material, take online classes; etc. 

These tasks usually spread through entire sessions.
Sometimes, one session embraces
multiple tasks. A quite common AA usage is to shout one or just a few
messages about current efforts, without much care for regularity or
completeness.

\subsection{Deviations from the AA paradigm}\label{sec:devia}
There are at least three perceived deviations from the idealized AA paradigm,
all most often within new users:
\begin{itemize}
    \item Advertising: shouts containing propaganda about events and groups.
	    This behavior is attributed to both 1) common practice in more commercial platforms,
		such as Twitter and Facebook;
		and 2) the outcome of the AA unusual goals and design,
		which requires acculturation before proper understanding.
    \item Final product exhibitionism: shouts containing not ongoing processes,
	    but only a media or deed recently completed.
		Although not considered entirely wrong by users,
		it does not accomplish the mechanism described in Section~\ref{sec:start}.
    \item Introduction to AA, IRC and hacking:
	    handling AA is regarded as empowerment and as an introduction to hacking and open co-working.
		The first contacts with AA is often marked by playful and test shouts.
		Although very well esteemed, these messages are also deviations from the AA purpose.
\end{itemize}

\section{Software support}\label{sec:sofsup}
There are mainly three software pieces written to support AA activity.
Two of them are a server and client suite each (see Sections~\ref{sec:aaFirst} and~\ref{sec:aa01}).
The third is a fancy dashboard.
Among supplementary software support are automated conversational agents (software (ro)bots),
used as alternative user interfaces (UIs),
with a highlight for the Lalenia bot (see Section~\ref{sec:lalenia});
and an initiative to make AA available in all chat networks (see Section~\ref{sec:ubi}).
All these software items are contextualized in Table~\ref{tab:aas}.

\begin{table}[H]
	\scriptsize
  \caption{All AA versions described and their databases.
	References marked with ($\dagger$) are not currently used.
	Further context is given in Sections~\ref{sec:sofsup} and~\ref{sec:data}.}\label{tab:aas}
\vspace{12pt}
\centering{}
  \begin{tabular*}{\textwidth}{@{\extracolsep{\fill}}|l|p{2cm}|p{2cm}|l|p{2.3cm}|l|}\hline
      {\bf version name} & {\bf main languages} & {\bf user interface} & {\bf database} & {\bf code repository} & {\bf available at} \\\hline\hline
First AA ($\dagger$) & PHP, Python, Bash & Linux terminal, HTML & MySQL & \cite{aafc,aafs} & -//- \\\hline
AA 0.1 & Python & Linux terminal, HTTP & MongoDB & \cite{aa01r} & \cite{aa01c,aa01s} \\\hline
Lalenia bot & Python & IRC & any & \cite{lalenia} & \cite{lirc} \\\hline
Ubiquitous AA ($\dagger$) & Python & IRC, Twitter, Google Chat, Facebook, email, MSN & any & \cite{ubi} & -//- \\\hline
  \end{tabular*}
\end{table}

\subsection{First AA: HTTP server, HTML skin and shell client}\label{sec:aaFirst}
Although deprecated in favor of AA 01,
this first AA software presents the most numerous set of functionalities.
The client was completely designed for usage within a GNU/Linux terminal,
and the main functionalities are:
\begin{itemize}
    \item Sending messages to the host.
    \item Configuration facilities.
    \item Access to tickets and other software development utilities.
    \item Timing to ease AA usage, as described in Sessions~\ref{sec:des} and~\ref{sec:usess}.
\end{itemize}

\noindent Main server functionalities are:
\begin{itemize}
    \item Receiving shouts and other AA messages through HTTP.
    \item Registering shouts and other AA messages, received through HTTP, in a MySQL database.
\end{itemize}

\noindent Core HTML interface functionalities are:
\begin{itemize}
    \item Exhibiting shouts and sessions to other users by common HTML pages.
    \item Enabling interaction of AA users for reviews and screencast attachments to sessions.
\end{itemize}

A full description of the features extrapolates the information above and the scope of this article,
as do implementation details.
Further information of this and other versions of AA are contextualized in Table~\ref{tab:aas} and~\cite{paaper}.

Throughout Jul/2011-Mar/2014, this first version of AA 
was used directly or routed from bots (see Section~\ref{sec:lalenia})
and other gadgets (Section~\ref{sec:ubi}).

\subsection{AA 0.1}\label{sec:aa01}
Although there was no online AA software support in the months of April and May, 2014,
there was AA activity, as seen in~\cite{ensaaio}.
This motivated the creation of a minimum version of AA to support this visceral usage.
This minimum version is AA 0.1~\citep{aa01r}.
This implementation targeted registering the shout messages independently.
All other characteristics were left to data mining and tagging performed by the users.
The minimum client has only one feature:
a simple HTML call. Integrated trivially as bash commands by scripts and bots.
The minimum server has more features:
\begin{itemize}
    \item it receives the shout with an associated nick and registers it to a MongoDB instance with the time stamp in which the message arrived.
    \item Returns all shouts as a string or as JSON.
    \item It has a Heroku Flask app, integrated to an online MongoDB.
	    All these are free online services. 
\end{itemize}
\noindent The minimum interface/skin features are part of the server, but listed here for organization:
\begin{itemize}
    \item Lightest HTML.
    \item Export as JSON.
\end{itemize}

\subsection{pAAinel}\label{sec:aaPaainel}
A fancy skin for visualizing AA activity is pAAinel.
Core features are visualization of:
\begin{itemize}
    \item Latest AA shouts.
    \item Latest IRC messages.
    \item Embedded Black Duck Open Hub (former Ohloh) analytics.
    \item Latest commits to LM main repositories.
\end{itemize}

\noindent This Django (Python) software was written for first AA and has not been adapted to AA 0.1.

\subsection{Lalenia interface}\label{sec:lalenia}
To ease and enhance the usage and social aspects of AA, the lalenia IRC bot was used as an AA client.
This enabled shouts to be logged by IRC users while on the same channel as lalenia.
Core features are:
\begin{itemize}
    \item Users on the same channel as lalenia can log shouts by using the prefix ``\texttt{;aa }'' to a regular message.
    \item Returns confirmation that the shout was logged.
	    Returns information about AA and software and concepts if successfully logged.
		Returns an error message if message was not logged.
    \item Both first version and 0.1 have supybot plug-ins (lalenia is a supybot~\citep{supybot}).
\end{itemize}

Within information in lalenia logs\footnote{About
the \#labmacambira@Freenode IRC channel log,
as stated by lalenia, {\tt on \#labmacambira there have been 172554 messages, containing 6964778 characters, 1066138 words, 4202 smileys, and 10181 frowns; 178 of those messages were ACTIONs.There have been 31906 joins, 680 parts, 31109 quits, 0 kicks, 4 mode changes, and 133 topic\\ changes}.},
there were found 1,654 AA shouts that were not in either MySQL or MongoDB databases.
Actually, to ease mining, any shout in channel log whose message is identical to any message in all 114,040 messages from 
MongoDB or MySQL was discarded.
Therefore, there was probably a few more shouts in \#labmarambira IRC channel log then what is reported here.

\subsection{Ubiquitous AA}\label{sec:ubi}
Following the route enabled by AA social network interfaces (Section~\ref{sec:lalenia} reports an example for IRC),
the Ubiquitous AA is the expansion of AA to be usable in all social networks and, indeed,
to any media in which activity can be registered.
There are two approaches to ubiquitous AA:
\begin{enumerate}
    \item a software that connects to many messaging services as a bot.
	    One implementation connected to all IRC, Twitter, Google Chat, Facebook, email and MSN.
		This bot usually receives messages and register them as AA shouts,
		but can encompass more elaborate communication procedures~\citep{ubi}.
    \item Tags with which messages are bind to AA activity.
	    This automatically enables AA usage in all social platforms.
		Messages can be mined for reports and other community usage.
\end{enumerate}

Ubiquitous AA has mythological aspects:
it is understood as the receiver part of the Yupana Kernel,
a mythological entity that receives and spreads all things~\citep{yupana};
and is also understood as a necessary step to human unification~\citep{wisaa,ciberiun}.

\section{Data}\label{sec:data}
AA data is scattered among different databases and logs (see Section~\ref{sec:sofsup}).
A coherent integration of these resources is done by means of a dedicated OWL ontology
and a mapping routine from relational and NoSQL databases to RDF data.

\subsection{The OntologiAA OWL ontology}\label{sec:ont}
An ontology, in linked data contexts, is a formalized conceptualization.
The Web Ontology Language (OWL) is a family of languages designed for authoring ontologies.
Core uses of ontologies include: 1) reasoning by means of ontological specifications;
2) linking data from different sources; and 3) organization of domain knowledge for coherent consideration.
For further ontologies in contexts pertinent to AA, the reader should visit~\cite{ops,pnud5}.

\begin{figure}[!htbp] %h or !htbp
\vspace{-2pt}
\begin{center}
    \includegraphics[width=0.7\textwidth]{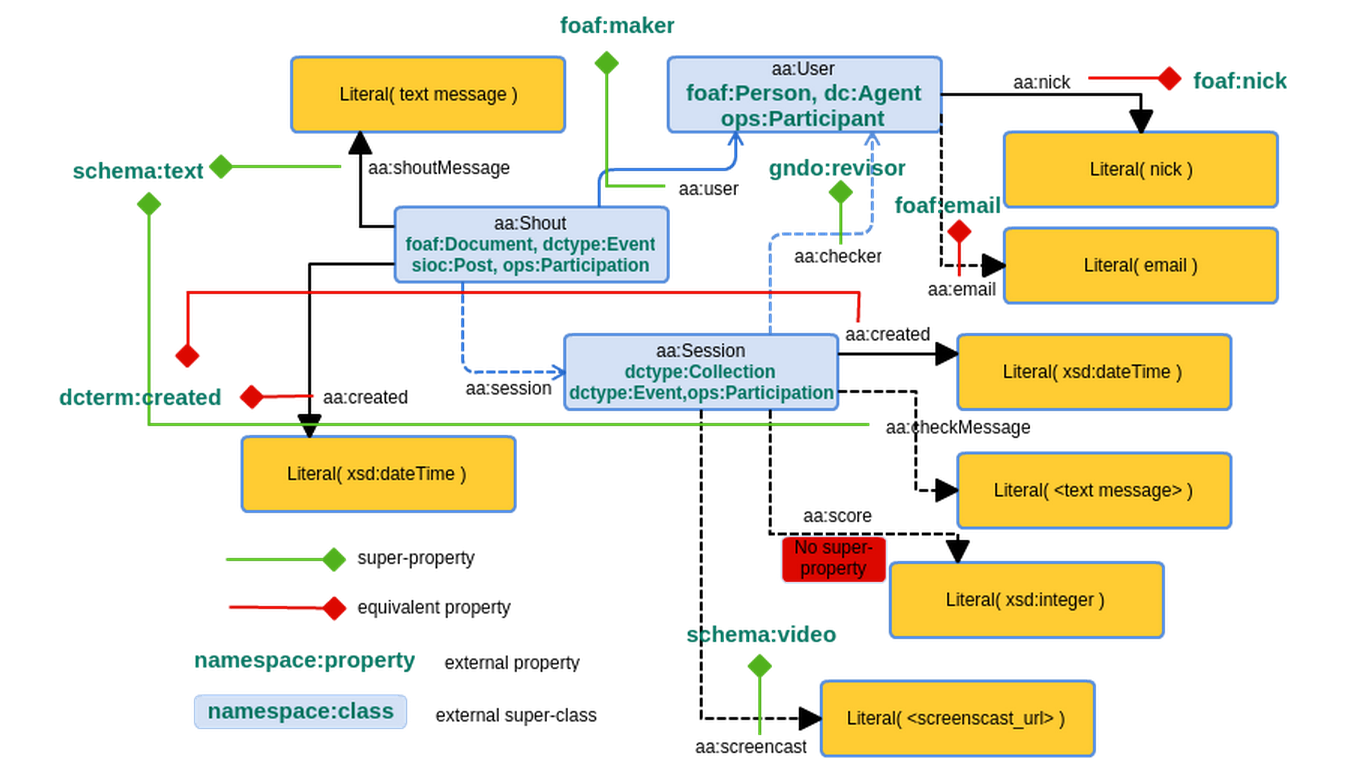}
    \caption{The OntologiAA: an OWL ontology of AA. Classes are concepts related by properties.
	Classes are also related to upper ontologies (FOAF, Dublin Core, Schema, SIOC, GNDO, OPS), as are the properties.
	All properties are functional, except for {\tt aa:nick} and {\tt aa:email}.
	Properties with a full line yield existential restrictions to the subjects of the triples suggested by the diagram: 
	{\tt aa:user}, {\tt aa:nick}, {\tt aa:shoutMessage}, {\tt aa:created}.
	Blue lines mark object properties, while black lines are for data properties.
	For further information, see Section~\ref{sec:ont}.}.
    \label{fig:ontologiaa}
\end{center}
\end{figure}

For AA, an ontology facilitates key usage aspects:
\begin{itemize}
    \item The conceptualization of AA is not unique or steady nor always easy to grasp.
	    The OntologiAA delivers a formal paradigm with which the community can communicate and develop.
		Figure~\ref{fig:ontologiaa} illustrates the conceptualization formalized in OntologiAA.
    \item There is AA related data in different databases,
	    and even in social networks (see Section~\ref{sec:sofsup}).
		Standard classes to which relate data is a sound method for integration.
    \item AA data is integrated to other social participation instances,
	    such as~\cite{participa,cd}, by super-classes and super-properties of the OntologiAA.
\end{itemize}

\subsection{RDF data}
The AA data is currently in relational and NoSQL databases,
and social networks to be mined (Section~\ref{sec:sofsup}).
By using the same ontological background (see OntologiAA in Section~\ref{sec:ont}),
this data can be translated to RDF for integration and linkage.

The scripts at~\cite{participation} output RDF from a MySQL database (mostly from first AA version),
from a MongoDB database (mostly from AA 0.1),
and from IRC logs.

\subsection{Linkage to external data}
The usage of RDF and OWL protocols enables linked data facilities.
The OntologiAA (and thus all AA data) is integrated by two means of special interest:
\begin{itemize}
    \item Through OPS, AA data is linked to OPa, OBS, VBS, and OCD.
	    These are ontologies
		with a social participation focus
		that have been used for data representation and conceptual studies~\citep{pnud5}.
    \item Through all other ontologies (e.g. FOAC, Dublin Core, Schema.org, SIOC),
	    AA data is linked to the Giant Global Graph (GGG) of Linked Open Data (LOD)~\citep{LOD,losd}.
\end{itemize}

This linkage gives meaning to data while facilitating data discovery and comparative analysis,
to point just a few of the benefits of such approach~\citep{ldb}.

\section{Statistics}\label{sec:stats}
From all the AA data, many statistics were obtained:
number of messages by type, of sessions, of users, screescants, sessions checked, scores, etc.;
activity along time in different scales from seconds to years;
most incident words, radicals and sizes of different sets of tokens.
All these and some other measures are available in~\cite{ensaaio}.
Most importantly, in the histograms it became clear that even in the months when no
software were being used to log the shouts, AA kept being used in IRC
within the Ubiquitous AA paradigm described in Section~\ref{sec:ubi}.

\section{CONCLUSIONS AND FURTHER WORK}\label{sec:conc}
This article describes effectively the AA design goals, the software implementations
it received and the usage entailed.
A preliminary thorough exposition of this content together with the statistical measures
reached 29 pages~\citep{ensaaio} and is better presented herein where the thorough analysis
is an auxiliary document.

Potential next steps are:
\begin{itemize}
	\item implementing mining routines or chat bots to better support the Ubiquitous AA
		as described in Section~\ref{sec:ubi}.
	\item Making the data related to AA available in the linked open data cloud~\citep{LOD,losd},
		e.g. through uploading OntologiAA and data to Data Wold or DataHub.
	\item Deepening our understandings of related paradigms, such as Pomodoro~\citep{pomodoro}.
	\item Relating the data analysis in~\cite{ensaaio} to data from other online social platforms.
\end{itemize}

\subsection*{\textit{Acknowledgements}}
The author thanks CNPq for the funding received while researching the topic of this article,
the researchers of IFSC/USP and ICMC/USP for the recurrent collaboration in every situation
where we needed directions for investigation.
The author also thanks the labMacambira.sf.net members for conceptualizing and using AA
and developing the software support.

% ------------------------------------------------------------------------

\end{document}